\documentclass[prl,reprint,showpacs,amsmath,amssymb,superscriptaddress,longbibliography]{revtex4-1}
\usepackage{graphicx,bm,color,mathptmx,hyperref} 
\newcommand{\Tr}{\mathop{\mathrm{Tr}} \nolimits}
\newcommand{\op}[1]{\hat{#1}}

\begin{document}
\title{Parsing Squeezed Light into Polarization Manifolds}

\author{C.~R.~M\"{u}ller} 
\affiliation{Max-Planck-Institut f\"ur die Physik des Lichts, 
  G\"{u}nther-Scharowsky-Stra{\ss}e 1, Bau 24, 91058 Erlangen, Germany} 
\affiliation{ Institut f\"ur Optik, Information und Photonik, 
 Universit\"{a}t Erlangen-N\"{u}rnberg,  Staudtstra{\ss}e 7/B2, 
 91058 Erlangen, Germany}
\affiliation{Department of Physics, Technical  University of Denmark, 
Fysikvej, 2800 Kgs. Lyngby, Denmark}

\author{L.~S.~Madsen} 
\affiliation{Department of Physics, Technical  University of Denmark, 
Fysikvej, 2800 Kgs. Lyngby, Denmark}
	
\author{A.~Klimov} 
\affiliation{Departamento de F\'{\i}sica,
  Universidad de Guadalajara, 44420~Guadalajara, Jalisco, Mexico}
	  
\author{L.~L.~S\'anchez-Soto} 
\affiliation{Departamento de \'Optica,
  Facultad de F\'{\i}sica, Universidad Complutense, 
  28040~Madrid, Spain} 
\affiliation{Max-Planck-Institut f\"ur die Physik des Lichts, 
  G\"{u}nther-Scharowsky-Stra{\ss}e 1, Bau 24, 91058 Erlangen, Germany} 
\affiliation{ Institut f\"ur Optik, Information und Photonik, 
 Universit\"{a}t Erlangen-N\"{u}rnberg,  Staudtstra{\ss}e 7/B2, 
 91058 Erlangen, Germany}

\author{G.~Leuchs} 
\affiliation{Max-Planck-Institut f\"ur die Physik des Lichts, 
  G\"{u}nther-Scharowsky-Stra{\ss}e 1, Bau 24, 91058 Erlangen, Germany} 
\affiliation{ Institut f\"ur Optik, Information und Photonik, 
 Universit\"{a}t Erlangen-N\"{u}rnberg,  Staudtstra{\ss}e 7/B2, 
 91058 Erlangen, Germany}

\author{Ch.~Marquardt} 
\affiliation{Max-Planck-Institut f\"ur die Physik des Lichts, 
  G\"{u}nther-Scharowsky-Stra{\ss}e 1, Bau 24, 91058 Erlangen, Germany} 
\affiliation{ Institut f\"ur Optik, Information und Photonik, 
 Universit\"{a}t Erlangen-N\"{u}rnberg,  Staudtstra{\ss}e 7/B2, 
 91058 Erlangen, Germany}
\affiliation{Department of Physics, Technical  University of Denmark, 
Fysikvej, 2800 Kgs. Lyngby, Denmark}

\author{U.~L.~Andersen} 
\affiliation{Department of Physics, Technical  University of Denmark, 
Fysikvej, 2800 Kgs. Lyngby, Denmark}
\affiliation{Max-Planck-Institut f\"ur die Physik des Lichts, 
  G\"{u}nther-Scharowsky-Stra{\ss}e 1, Bau 24, 91058 Erlangen, Germany} 
\email{ulrik.andersen@fysik.dtu.dk}

\begin{abstract}
  We investigate polarization squeezing in squeezed coherent states
  with varying coherent amplitudes. In contrast to the traditional
  characterization based on the full Stokes parameters, we
  experimentally determine the Stokes vector of each excitation
  manifold separately. Only for states with a fixed photon number do
  the methods coincide; when the photon number is indefinite, our
  approach gives a richer and deeper description. By capitalizing on
  the properties of the Husimi $Q$ function, we map this notion onto
  the Poincar\'e space, providing a full account of the measured
  squeezing.
\end{abstract}

\pacs{42.50.Dv, 42.50.Lc, 03.65.Wj}

\maketitle

Heisenberg's uncertainty principle~\cite{Busch:2007fk} epitomizes the
basic tenets of quantum theory and it comes out as a strict trade-off:
fluctuations of a given observable can always be reduced below some
threshold at the expense of an increase in the fluctuations of another
observable.  A time-honored example of this trade-off is provided by
quadrature squeezed states of light~\cite{Lvovsky:2015qv}, which can
be generated, for example, with lower uncertainty in their amplitude
and higher uncertainty in their phase.

The notion of squeezing, while universal for harmonic oscillator-like
systems, is otherwise far from unique. For spin-like systems there are
several approaches~\cite{Kitagawa:1993wb,Itano:1993fk,Ma:2011xd}: all
of them compare fluctuations of suitably chosen observables with a
threshold given by some reference state. Spin squeezed states have
attracted a lot of attention in recent years as they might constitute
an important resource in quantum
information~\cite{Gross:2012fk,Guhne:2009fk}.

As the Stokes operators~\cite{Jauch:1955fj}, specifying the
polarization properties of quantum fields, match the standard features
of an angular momentum, the parallel between spin and polarization
squeezing~\cite{Chirkin:1993dz} cannot come as a surprise. Actually,
for states with fixed photon number both notions coincide and have
been experimentally demonstrated~\cite{Shalm:2009gd}.  In the opposite
regime of an undefinite number of photons (often involving bright
states~\cite{Korolkova:2002kb}), polarization squeezing has been
reported in numerous systems, including parametric
amplifiers~\cite{Lassen:2007th,Iskhakov:2009by}, optical
fibers~\cite{Heersink:2005ul,Marquardt:2007bh}, and atomic
vapors~\cite{Josse:2003ca,Barreiro:2011rf}. The uncertainty in photon
number now forces us to scrutinize multiple excitation manifolds.

Until now there have been no studies on the transition between these
two regimes. The goal of this Letter is to explore both within
a single experiment. Using two optical parametric amplifiers,
complemented with a phase-space displacement, we squeeze various fixed
photon-number manifolds. Polarization squeezing is analyzed as a
function of the coherent amplitude, finding out that this operation
tends to degrade squeezing. Besides, this transition from vacuum
squeezing to displaced vacuum squeezing can be clearly visualized in
Poincar\'{e} space using the appropriate Husimi $Q$ representation.

Let us start by briefly recalling some basic notions. We shall be
dealing with monochromatic fields, defined by two operators
$\op{a}_{H}$ and $\op{a}_{V}$: they represent the complex amplitudes
in two linearly polarized orthogonal modes, that we indicate as
horizontal ($H$) and vertical ($V$), respectively. The Stokes
operators are~\cite{Luis:2000ys}
\begin{equation}
  \label{eq:Stokop}
  \begin{array}{c}
    \op{S}_{x} = \textstyle\frac{1}{2} 
    ( \op{a}^{\dagger}_{H}  \op{a}_{V} + 
    \op{a}^{\dagger}_{V} \op{a}_{H} ) \, ,  
    \qquad
    \op{S}_{y} =  \frac{i}{2} ( \op{a}_{H} \op{a}^{\dagger}_{V} - 
    \op{a}^{\dagger}_{H} \op{a}_{V} ) \, ,  \\
    \\
    \op{S}_{z}  = \frac{1}{2} ( \op{a}^{\dagger}_{H} \op{a}_{H} - 
    \op{a}^{\dagger}_{V} \op{a}_{V} ) \, ,
  \end{array}
\end{equation}
together with the total photon number
$ \op{N} = \op{a}^{\dagger}_{H} \op{a}_{H} + \op{a}^{\dagger}_{V}
\op{a}_{V}$.
The components of the vector
$\op{\mathbf{S}} = (\op{S}_{x}, \op{S}_{y}, \op{S}_{z})$ thus satisfy
the commutation relations of the su(2) algebra:
$[ \op{S}_{x}, \op{S}_{y}] = i \op{S}_{z}$ and cyclic permutations (we
use $\hbar =1$ throughout).

In classical optics, we have a Poincar\'e sphere with radius equal to
the intensity, which is a sharp quantity.  In contradistinction, in
quantum optics Eq.~\eqref{eq:Stokop} implies that
$\op{\mathbf{S}}^{2} = \op{S}_{x}^{2} + \op{S}_{y}^{2} +
\op{S}_{z}^{2} = S (S + 1) \op{\openone}$,
with $S =N/2$ playing the role of the spin.  When the photon number is
fuzzy, we need to consider a three-dimensional Poincar\'e space (with
axis $S_{x}$, $S_{y}$, and $S_{z}$). This space can be visualized as a
set of nested spheres with radii proportional to the diverse photon
numbers that contribute to the state and that can be aptly called the
polarization manifolds.

Since $ [ \op{N}, \op{\mathbf{S}} ] = 0$, each excitation manifold
should be addressed independently.  This can be underlined if instead
of the Fock basis $\{ |n_H, n_V \rangle \} $, we employ the relabeling
$ | S, m \rangle \equiv | n_H = S + m, n_V = S - m \rangle$ that can
be seen as the common eigenstates of $ \op{S}^{2}$ and $\op{S}_{z}$.
Note that $S=(n_H+n_V)/2$ and $m = (n_H-n_V)/2$. Moreover, the moments
of any energy-preserving observable $f(\op{\mathbf{S}} )$ do not
depend on the coherences between manifolds or on global phases: the only
accessible polarization information from any density matrix
$\op{\varrho}$ (which describes the state) is in its block-diagonal form
$\op{\varrho}_{\mathrm{pol}} = \oplus_{S} \op{\varrho}^{(S)}$, where
$\op{\varrho}^{(S)}$ is the reduced density matrix in the subspace
with spin $S$. Accordingly, we drop henceforth the subscript pol.
This $\op{\varrho}_{\mathrm{pol}}$ has been dubbed the polarization
sector~\cite{Raymer:2000zt} or the polarization density
matrix~\cite{Karassiov:2004xw}.

\begin{figure}
  \includegraphics[width=0.90\columnwidth]{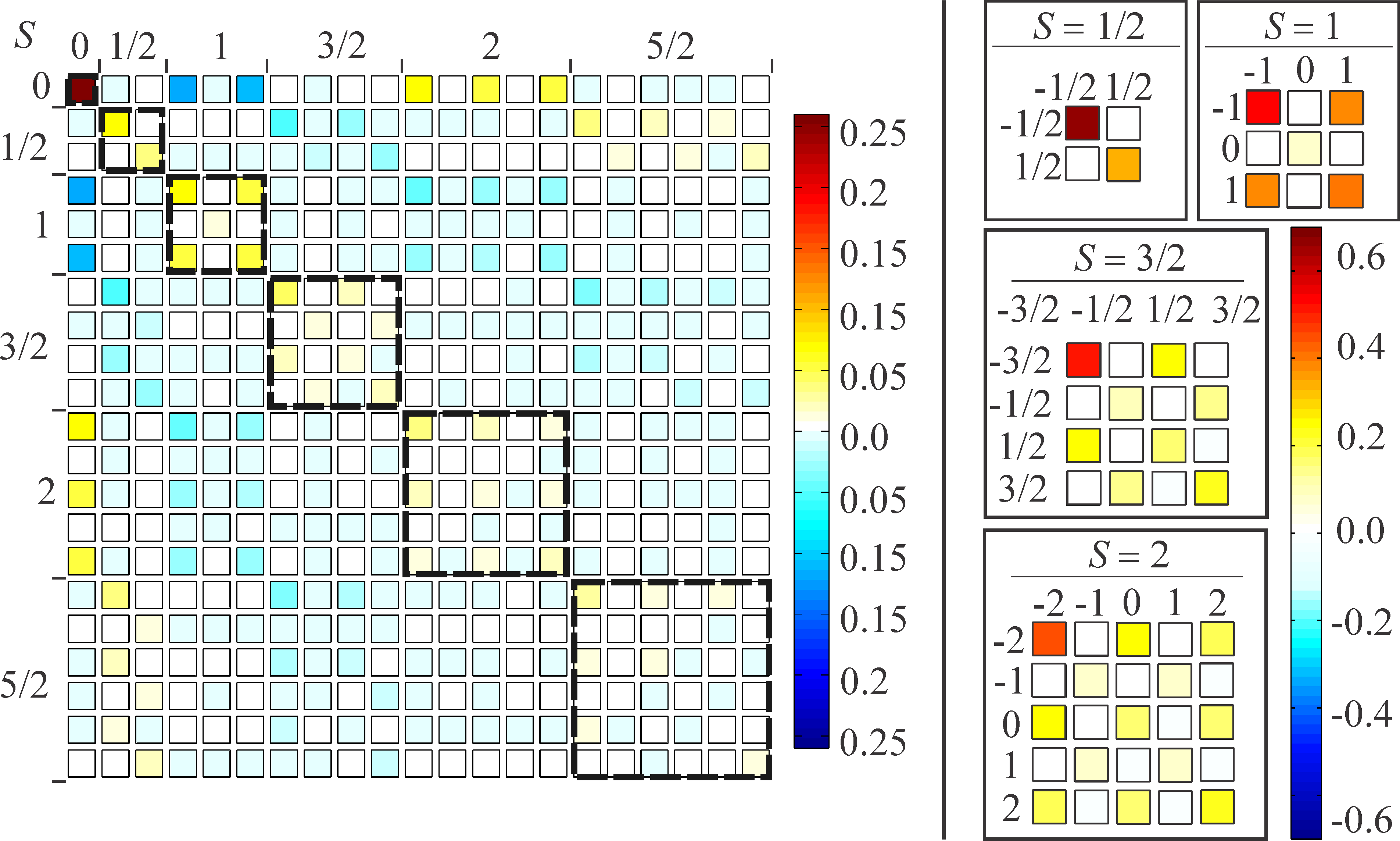}
  \caption{(color online) Illustration of the measured polarization
    sector (black blocks) for a polarization squeezed state as in
    Eq.~\eqref{eq:SqueezedStates}, with $\alpha = 1.13$.  The subplots
    at the right depict different individually-normalized manifolds.}
  \label{density_matrices}
\end{figure}

An example of the density matrix of
one of our experimentally acquired states is shown in
Fig.~\ref{density_matrices}, where the sub-matrices associated with
different polarization manifolds are displayed.

The shot-noise limit in the manifold of spin $S$ (i.e., $N=2S$
photons) is settled in terms of SU(2) (or spin) coherent
states~\cite{Perelomov:1986ly}. They are defined as
$|S, \mathbf{n} \rangle = \op{D} ( \mathbf{n} ) |S, S \rangle$, where
$\mathbf{n}$ is a unit vector [with spherical angles ($\theta , \phi$)]
on the Poincar\'e sphere of radius $\sqrt{S(S+1)}$ and $\op{D} 
( \mathbf{n} ) =  e^{i \theta \op{S}_{y}} e^{i \phi  \op{S}_{z}}$
plays the role  of a displacement  on that sphere.  For 
these states the  variances of the Stokes operators
[$\Delta^{2} \op{S}_{k} = \langle \op{S}_{k}^{2} \rangle - \langle
\op{S}_{k} \rangle^{2}$] depend on $\mathbf{n}$, and there exists 
a preferred direction: the mean-spin direction. The corresponding
variances  in the direction $\mathbf{n}_{\perp}$ perpendicular to the 
mean spin are isotropic and $\Delta^{2} \op{S}_{\mathbf{n}_{\perp}} =
S/2$,  which is taken as the shot noise. In consequence, polarization 
squeezing for an arbitrary state occurs whenever the condition
$\inf_{\mathbf{n}} \Delta^{2} \op{S}_{\mathbf{n}} < S/2$ holds true.

A way to get around the dependence on the directions is to use the
real symmetric $3 \times 3$ covariance matrix for the Stokes
variables~\cite{Klimov:2010uq}, defined as
$ \Gamma_{k \ell} = \frac{1}{2} \langle \{ \op{S}_{k}, \op{S}_{\ell}
\} \rangle - \langle \op{S}_{k} \rangle \langle \op{S}_{\ell}
\rangle$,
where $\{ , \}$ is the anticommutator. In terms of this matrix
$\Gamma$, we have
$ \Delta^{2} \op{S}_{\mathbf{n}} = \mathbf{n}^{t} \, \Gamma \,
\mathbf{n}$
and, since $\Gamma$ is positive definite, the minimum of
$ \Delta^{2} \op{S}_{\mathbf{n}}$ exists and it is unique. If we
incorporate the constraint $\mathbf{n}^{t} \cdot \mathbf{n} =1$ as a
Lagrange multiplier $\gamma$, this minimum is given by
$\Gamma \mathbf{n} = \gamma \mathbf{n}$: the admissible values of
$\gamma$ are thus the eigenvalues of $\Gamma$ and the directions
minimizing $ \Delta^{2} \op{S}_{\mathbf{n}}$ are the corresponding
eigenvectors. Therefore, we can define the degree of polarization
squeezing as
\begin{equation}
  \label{eq:polsqzdef}
  \xi^{2} = \inf_{\mathbf{n}} \frac{\Delta^{2}
    \op{S}_{\mathbf{n}}}{S/2} = \frac{4 \gamma_{\mathrm{min}}}{N}  \, .
\end{equation}
We stress, though, that this definition is not unique and a number of
proposals can be found in the literature, each one being specially
tailored for specific purposes~\cite{Ma:2011xd}.

When the state spans several manifolds, we follow
Ref.~\cite{Kothe:2013fk} and bring to bear an averaged Stokes vector
$ \langle \op{\mathbf{S}} \rangle = \sum_{S=0}^{\infty} P_{S} \; \Tr
(\op{\varrho}^{(S)} \op{\mathbf{S}} )$,
where $P_{S}$ is the photon-number distribution. As a result, the
squeezing of the state can be much lower than the corresponding one in
the individual manifolds.

\begin{figure}
  \includegraphics[width=.90\columnwidth]{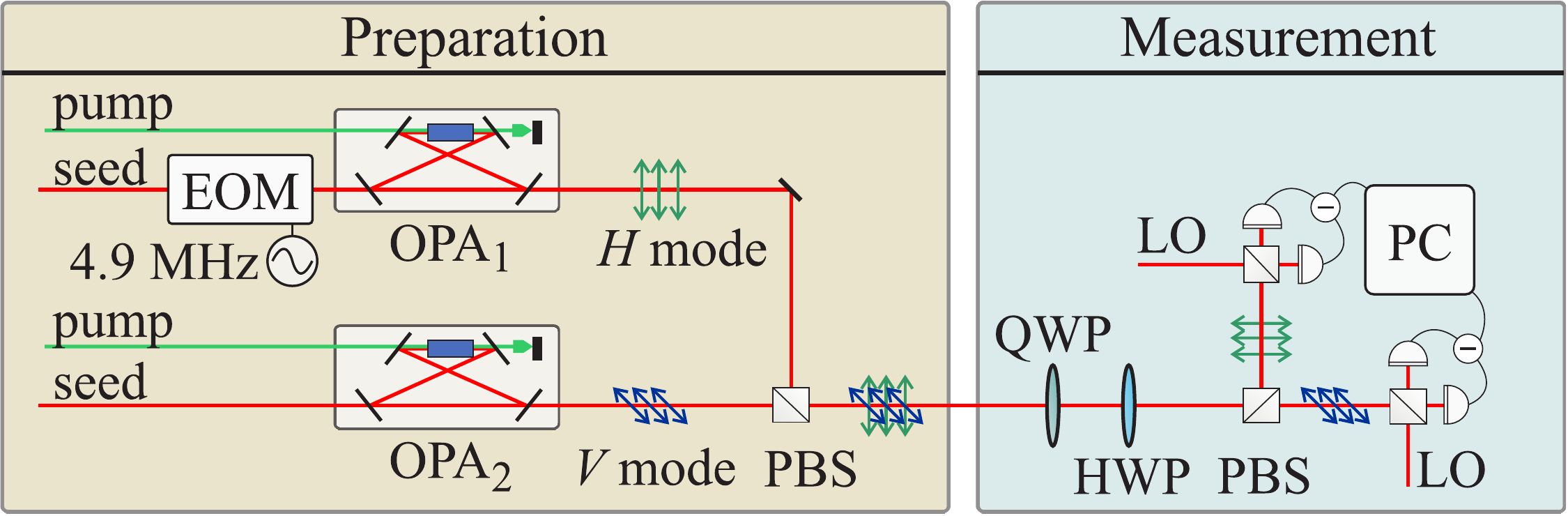}
  \caption{(color online) Experimental setup. Two optical parametric
    amplifiers (OPA1 and OPA2) independently squeeze coherent seed
    beams in orthogonal polarization modes $H$ and $V$.  The seed beam
    entering OPA1 is modulated at the sideband frequency of 4.9~MHz.
    The modes are spatially combined on a polarizing beam splitter
    (PBS) and interference between the modes can be adjusted with the
    combination of a quarter wave-plate (QWP) and a half wave-plate
    (HWP). The polarization states are separated into orthogonal
    components followed by homodyne tomography.}
  \label{fig:Setup}
\end{figure}

To confirm these issues we use the setup sketched in
Fig.~\ref{fig:Setup}.  It comprises two optical parametric amplifiers
(OPA1 and OPA2) operating below threshold and pumped with a 532~nm
continuous-wave laser beam to produce two quadrature squeezed states.
The parametric down-conversion processes are based on type I
quasi-phase-matched periodically poled KTP crystals and generate
squeezed states in one polarization mode.  The OPAs were seeded with
dim laser beams at 1064~nm to facilitate the locking
(Pound-Drever-Hall technique~\cite{Drever:1983ty}) of the cavities and
several phases of the experiment.  One of the seed beams is modulated
via an electro-optical modulator (EOM) at the sideband frequency of
4.9~MHz relative to the carrier frequency and with variable modulation
depth, allowing to control the amplitude of the thereby generated
coherent states.  The resulting modes are combined on a polarizing
beam splitter (PBS) to form the state
\begin{equation}
  | \Psi \rangle  = \op{S} (r_{H} ) \op{D} (\alpha_{H}) | 0_{H}\rangle
  \otimes \op{S} (r_{V} ) | 0_{V} \rangle \, .
  \label{eq:SqueezedStates}
\end{equation}
Here, $\op{D}(\alpha) = \exp( \alpha \op{a}^{\dagger} - \alpha^\ast \op{a})$
is the displacement and $\op{S} (r)= \exp [ (r^\ast \op{a}^2 - 
r \op{a}^{\dagger 2})/2]$ is the squeezing operator. The indexes $H$
and $V$ denote the mode to which the operator is applied.  Since
$r_{H} \simeq r_{V} \simeq 0.41$, we drop the corresponding
subscripts. In addition, $\alpha$ will subsequently designate the
amplitude of the horizontal component of the state after OPA1, which
has been constrained to be a real number. Experimentally, we achieve
about 3.6~dB of quadrature squeezing in both modes and about 4.4~dB of
excess noise along the anti-squeezing direction.

To characterize the polarization state, the beam is directed to the
verification stage.  A quarter wave-plate (QWP) and a half-wave plate
(HWP) allow to verify the interference between the orthogonal
polarization modes and to control the measurement basis.  To ease the
otherwise complicated two-mode tomography, the original horizontal and
vertical polarization modes are separated by a PBS and each one is
characterized via homodyne tomography.  The measurement is performed
at the sideband frequency of 4.9~MHz with a bandwidth of 90~kHz.  The
local oscillator (LO) phases are scanned continuously to acquire the
tomographic data and the homodyne outputs are stored in a computer. We
determine the phase of the LO scans in the data, which allows one to
compensate for any phase drifts among the two measurement stages in
the post processing steps. The same data are then analyzed for noise
properties at the measurement frequency.

\begin{figure}
  \begin{center}
    \includegraphics[width=1.03\columnwidth]{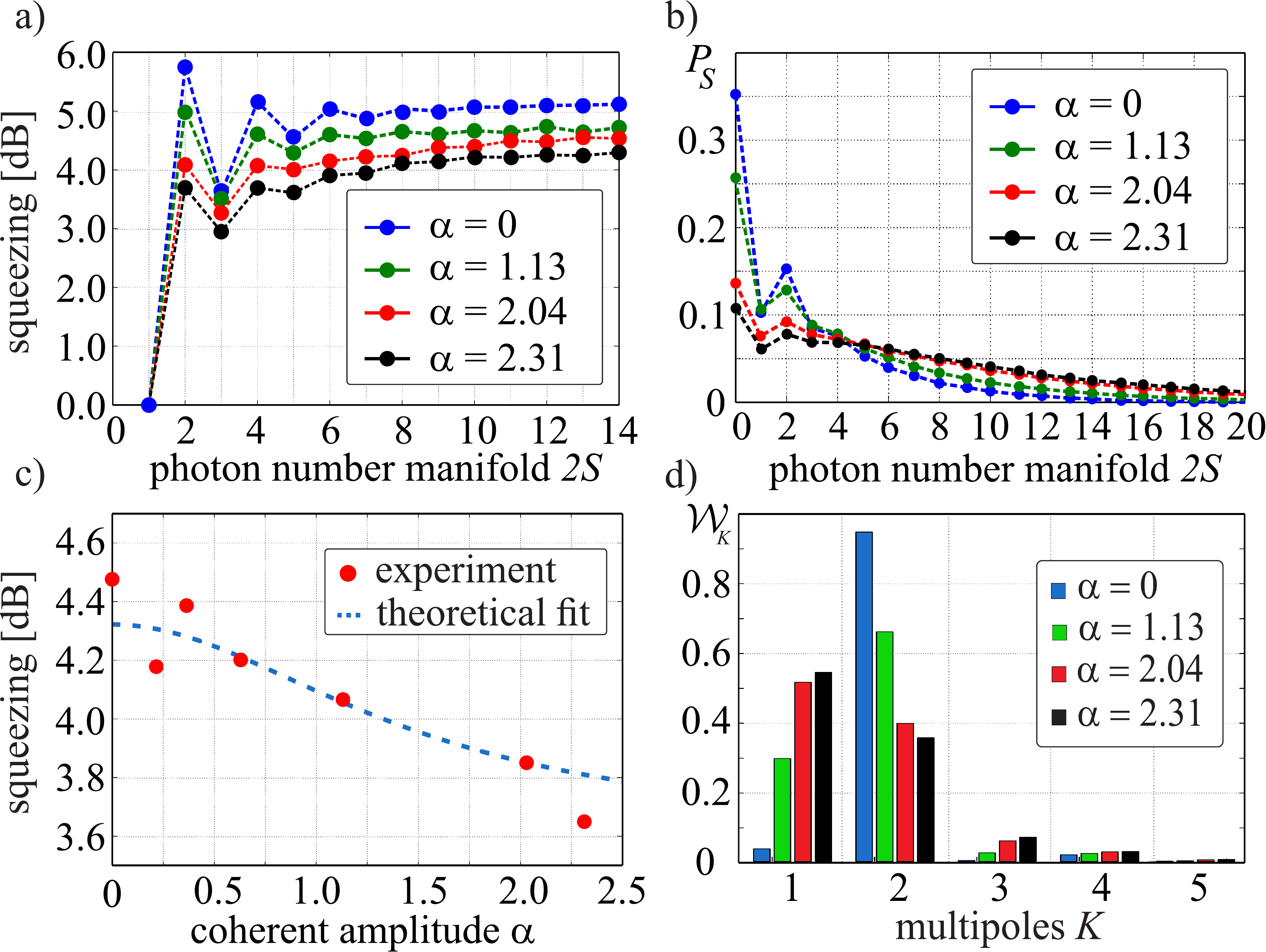}
  \end{center}
  \caption{(color online) Experimental results for different values of
    the coherent amplitude $\alpha$.  (a) Polarization squeezing as a
    function of the excitation manifold.  (b) Photon-number
    distributions.  (c) Polarization squeezing of the total state as a
    function of the coherent amplitude $\alpha$.  (d) Distribution of
    $\mathcal{W}_{K}$ as a function of the multipole order $K$.}
  \label{fig:Squeezing}
\end{figure}

In Fig.~\ref{fig:Squeezing}(a) we plot the measured polarization
squeezing in the different excitation manifolds for various values of
the coherent amplitude $\alpha$.  Squeezing
occurs for all photon numbers except for the vacuum and the
one-photon manifolds.  This is intuitively clear, for squeezing the
Stokes variables involves nonclassical correlations among individual
photons: such correlations thus require the presence of at least two
photons.  For $S=1$, these correlations are dramatically demonstrated
by the presence of 6~dB polarization squeezing (for $\alpha=0$).

Polarization squeezing is not equally distributed among the manifolds,
but exhibits an oscillating pattern that is most pronounced for small
S and small amplitude states.  If the individual modes were ideally
squeezed vacua ($\alpha = 0$) and were measured with perfect
detectors, only even-photon number manifolds would contribute.  Due to
the additional excess noise of about 4.4~dB and the finite efficiency
of the homodyne detectors ($98\%$ quantum efficiency of the photo
diodes, $85\pm5\%$ total efficiency) , however, the photon-number
contributions are smeared out, as corroborated by Fig. 3(b), and
polarization squeezing can also be observed for odd-excitation
manifolds.  

Increasing $\alpha$ results in an overall reduction of the
polarization squeezing.  The coherent amplitude acts much the same as
a local oscillator, in such a way that the spin squeezing is
continuously transferred into a quadrature
measurement~\cite{Wang:2003qd}. This coincides with the direct
measurement of the bright squeezed vacuum states in
Ref.~\cite{Muller:2012fk}

The polarization squeezing of the entire state is also presented in
Fig.~\ref{fig:Squeezing}(c) as a function of $\alpha$.  The
experimental results are compared to a numerical simulation based on
the measured single-mode squeezing and excess noise.  For small
amplitudes, mainly the inner manifolds dominate the squeezing. In the
opposite limit of large amplitudes, the Stokes measurement reduces to a
quadrature measurement, and thus the degree of polarization squeezing
will no longer be determined by the photon-number correlations but by
quadrature correlations. Deviations from the theoretical curve are due
to small fluctuations in the squeezing and excess noise parameters
between individual measurement runs.

\begin{figure*}
  \centering
  \includegraphics[width=1.90\columnwidth]{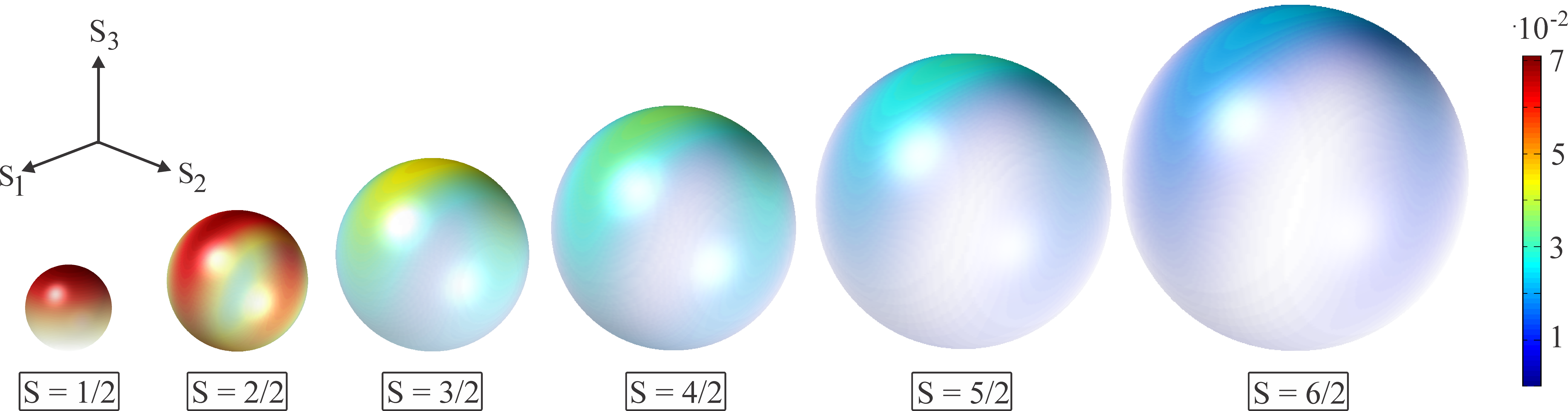}
  \caption{(color online) Reconstructed SU(2) $Q^{(S)}$ functions of the
    excitation manifolds indicated in the insets for a polarization
    squeezed state with $\alpha = 1.13$. The scale of the density
    plots on the corresponding Poincar\'e spheres is shown on the
    right.}
  \label{fig:Qs}
\end{figure*}

It is worth stressing that parsing the state into manifolds turns out
to be crucial to analyze the experimental results. If one computes the
covariance matrix of Eq.~\eqref{eq:SqueezedStates} deemed as a
two-mode state, one gets, after some calculations,
\begin{equation}
  \label{eq:var}
  \xi^{2}  = \frac{4 \gamma_{\mathrm{min}}}{\langle \op{N} \rangle} 
  \simeq \frac{|\alpha|^{2} e^{-r}}
  {|\alpha|^{2} + \frac{1}{2} \sinh^{2} r} \, ,
\end{equation}
where a direct extension of Eq.~\eqref{eq:polsqzdef} has been used.
Whereas this gives the correct limit discussed before for
$\alpha \rightarrow \infty$, it fails to reproduce the observed
squeezing for $\alpha \rightarrow 0$.

We also lay out a phase-space picture of our previous discussion.  A
very handy way to convey the full information of the density matrix
$\rho^{(S)}$ associated to our states in Eq.~\eqref{eq:SqueezedStates} is
through the Husimi $Q$ function, defined as
$Q^{(S)} (\mathbf{n}) = \langle S, \mathbf{n} | \op{\varrho}^{(S)} |
S, \mathbf{n} \rangle$. In this way, $Q^{(S)} ( \mathbf{n} )$ appears 
as the projection onto SU(2) coherent states, which have the most 
definite polarization allowed by quantum theory. When the state
involves  multiple manifolds we have
\begin{equation}
  \label{eq:QSU2}
  Q ( \mathbf{n}  ) = \sum_{S} \frac{2S+1}{4 \pi}
  Q^{(S)} ( \mathbf{n} ) \, .
\end{equation}
This is an appealing feature of this function: because of the lack of
the off-diagonal contributions with $S \neq S^\prime$, the $Q$
function takes the form of an average over the manifolds with definite
total number of excitations.  Actually, the sum over $S$ in
Eq.~(\ref{eq:QSU2}) removes the total intensity of the field in such a
way that $Q ( \mathbf{n} )$ contains only the relevant polarization
information.

\begin{figure}[b]
  \centering
  \includegraphics[width=0.95\columnwidth]{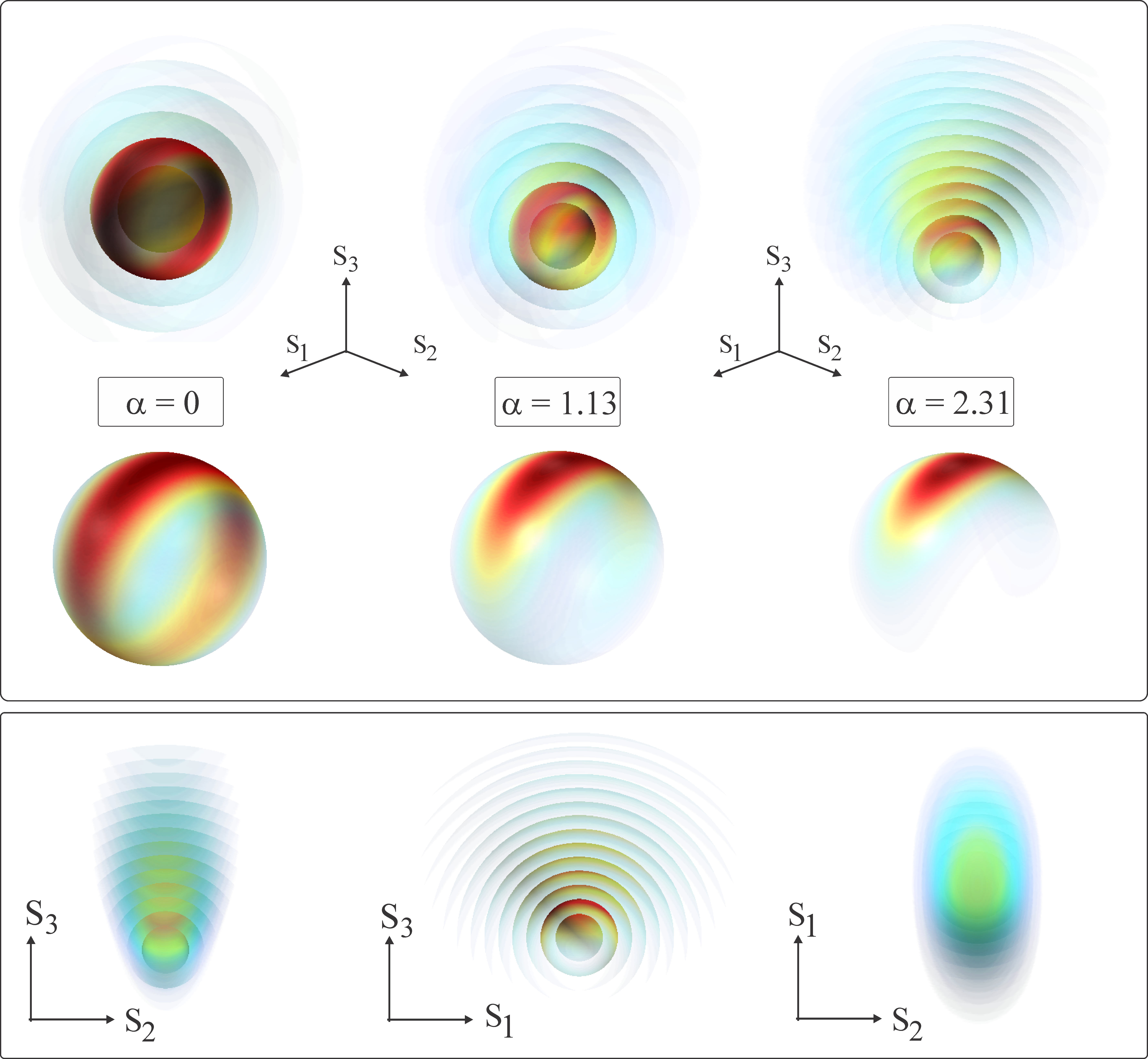}
  \caption{(color online) Top panel: Reconstructed SU(2) Husimi
    functions of the polarization squeezed states for three different
    coherent amplitudes $\alpha = 0, 1.13,$ and 2.31, from left to
    right. In the upper row, we represent the parsed version, foliated
    into suitably scaled polarization manifolds in Poincar\'e space.
    In the lower row, we have the total $Q$ function as given by
    Eq.~\eqref{eq:QSU2}.  Bottom panel: Views along the coordinate
    axes of the state with $\alpha=2.31$.}
  \label{fig:Qcompl}
\end{figure}

The expansion coefficients of $Q^{(S)}(\mathbf{n})$ in spherical
harmonics, which are a basis for the functions on the sphere
$\mathcal{S}^{2}$, read
\begin{equation}
  \varrho_{Kq}^{(S)} = \sqrt{\frac{2S+1}{4 \pi}} \frac{1}{C_{SS,K0}^{SS}}
  \int_{\mathcal{S}^{2}} d^{2}\mathbf{n} \ Y_{Kq} (\mathbf{n} ) \,
  Q^{(S)} (\mathbf{n}) \, ,
  \label{eq:Qtot} 
\end{equation}
where $K=0, \ldots , 2S$ and $C_{SS,K0}^{SS}$ is a Clebsch-Gordan
coefficient introduced for a proper normalization.  The
$\varrho_{Kq}^{(S)}$ are the standard state
multipoles~\cite{Blum:1981rb}, proportional to the $K$th power of the
Stokes variables. They can also be related to measures of state
localization on the sphere~\cite{Sanchez-Soto:2013cr}.

The quantity $ \mathcal{W}_{K}^{(S)} = \sum_{q=- K}^{K} |\varrho_{Kq}^{(S)}|^{2}$
is the square of the state overlapping with the $K$th multipole
pattern in the $S$th subspace.  When there is a distribution of photon
numbers, we sum over all of them to obtain
$\mathcal{W}_{K}$~\cite{Hoz:2013om}.  In Fig.~\ref{fig:Squeezing}(d)
we represent $\mathcal{W}_{K}$ as a function of the multipole order for
four values of the amplitude $\alpha$. From a practical
viewpoint only the dipole ($K=1$) and the quadrupole ($K=2$) are
noticeable. For $\alpha = 0$ the dipole is almost negligible while the
quadrupole is the leading contribution. The dipole becomes larger as
$\alpha$ increases, whereas the opposite happens for the quadrupole: a
clear indication that the state gets more and more localized.
 
In Fig.~\ref{fig:Qs} we plot the Husimi function of the first six
manifolds of a squeezed coherent state as in
Eq.~\eqref{eq:SqueezedStates}, with $ \alpha = 1.13$. The birth of
polarization squeezing is nicely observed: for the one-photon
manifold, the polarization spreads over the sphere and we expect no
squeezing, whereas in the two-photon manifold the uncertainty becomes
squeezed and belts around the sphere.  As the photon number is further
increased, the squeezing becomes more evident and the uncertainty area
becomes more localized, tracing out a squeezed ellipse on the sphere.

In Fig.~\ref{fig:Qcompl} the Husimi function of the entire state
parsed in its manifolds is illustrated for three displacements. When
$\alpha=0$, the innermost sphere with $S=1/2$ is highly occupied,
while the outer ones are almost empty.  A strong directional bias
appears when $\alpha$ increases. We also plot the total
$Q$ computed as in Eq.~\eqref{eq:QSU2}, wherein the squeezing becomes
conspicuous.  In the bottom panel we include the views of the parsed
Husimi function along the three coordinate axes for the state with
$\alpha = 2.31$. The typical cigar-like projections, familiar from
previous measurements~\cite{Muller:2012fk}, can be recognized.

In summary, we have presented a complete characterization of
polarization squeezing of squeezed coherent states. Parsing the
Poincar\'e space into excitation manifolds has played a pivotal role.
By varying the coherent amplitude, we have witnessed the transition
from states living in one single manifold to those spreading over many
of them. Far from being an academic curiosity, this has allowed us to
clarify previous discrepancies with the experiment.  Using the Husimi
$Q$ function for the problem at hand we have been able to envision
that transition in a very intuitive manner.

Financial support from the Lundbeck foundation, the Danish Council for
Independent Research (Sapere Aude 0602-01686B), the European Union FP7
(Grant QESSENCE), the European Research Council (Advanced Grant
PACART), the Mexican CONACyT (Grant 254127), and the Program
UCM-BSCH (Grant GR3/14) is gratefully acknowledged.


%

\end{document}